**Stressor-Layer-Induced Elastic Strain Sharing in SrTiO$_3$ Complex Oxide Sheets**

J. A. Tilka,[1] J. Park,[1] Y. Ahn,[1] A. Pateras,[1] Z. Cai,[2] and P. G. Evans[1*]

[1] Department of Materials Science and Engineering, University of Wisconsin-Madison,

Madison, Wisconsin 53706, USA

[2] Advanced Photon Source, Argonne National Laboratory, Argonne, Illinois 60439, USA

[*]email: pgevans@wisc.edu

**Abstract**

A precisely selected elastic strain can be introduced in submicron-thick single-crystal SrTiO$_3$ sheets using a silicon nitride stressor layer. A conformal stressor layer deposited using plasma-enhanced chemical vapor deposition produces an elastic strain in the sheet consistent with the magnitude of the nitride residual stress. Synchrotron x-ray nanodiffraction reveals that the strain introduced in the SrTiO$_3$ sheets is on the order of $10^{-4}$, matching the predictions of an elastic model. This approach to elastic strain sharing in complex oxides allows the strain to be selected within a wide and continuous range of values, an effect not achievable in heteroepitaxy on rigid substrates.

The deliberate introduction of elastic strain with a magnitude of on the order of 0.01% to 1% is a powerful way to control the functional properties of complex oxides. Elastic strain in nanoscale materials can raise or lower the mobility of charge carriers and ions, change the magnitude or symmetry of the electronic band gap, shift the Curie temperatures of magnetic and ferroelectric phase transitions, or modify ferroelectric nanodomain configurations through energy-competition mechanisms.[1-6] Physical phenomena arising from elastic strain are often studied and controlled in complex oxides by employing the large stresses arising from coherent heteroepitaxy, during which the in-plane lattice parameter of the thin film is constrained to match the lattice parameter of the substrate.[7] The use of this epitaxial lattice mismatch is, however, limited to a finite number of discrete values by the compositions of available substrates. For example, a $BaTiO_3$ (BTO) thin film coherently grown on $SrTiO_3$ (STO) has a fixed in-plane elastic strain of -2%.[8] Other values of the elastic mismatch strain can only be achieved by changing the composition, and hence lattice parameter, of the substrate.[9] It has not been possible to choose arbitrary values of the elastic strain because the range of compositions and lattice parameters of suitable single-crystal substrates is limited. In addition, a systematic exploration of strain effects can require the development of different surface preparation and epitaxial growth procedures for each substrate composition. We describe here an experimental test of an alternative approach in which a thin complex oxide sheet is strained by a stressor layer. The magnitude of the elastic strain is set by the magnitude of the stress, allowing the lattice parameter to be precisely selected. We demonstrate that this approach can be used to introduce strain in an STO sheet distorted by a calibrated silicon nitride layer without the formation of structural defects.

Elastic strain in complex oxides can be particularly important in ferroelectric and ferromagnetic materials. In ferroelectrics, the elastic strain, $\varepsilon$, in BTO grown on an elastically compliant STO sheet could in principle be set in the range of -2% to 0 by selecting the thicknesses of the BTO layer and STO sheet. Here strain is defined in terms of the unstressed and thin-film in-plane lattice parameters $a_{BTO}$ and $a_{BTO,film}$ as $\varepsilon = (a_{BTO,film} - a_{BTO})/a_{BTO}$. Varying the strain over this range would change the ferroelectric Curie temperature of BTO by more than 100 °C.[2] Similarly, the heteroepitaxy of the magnetic oxide $La_{0.7}Sr_{0.3}MnO_3$ on STO imposes a strain of on the order of 1% and relaxation near grain boundaries leads to regions of increased Curie temperature.[10]

Elastic strain sharing in oxides can employ processing techniques that have been developed for semiconductors, but which have not yet been explored more generally. The introduction of strain on the order of 1% has enabled dramatic improvements in the performance of Si field-effect transistors.[11] Elastic strain shifts the wavelength of infrared InGaAs heterostructure lasers grown on compliant substrates and can be used to induce a change in the bandgap of Ge from indirect to direct.[12,13,14] Strained-Si/relaxed SiGe heterostructures on compliant sheets exhibit reduced local variation in strain and tilt in comparison with layers grown on SiGe.[15] Advances in the control of strain in complex oxides promise to bring a similarly precise level of control.

A key challenge in elastic strain sharing in complex oxides has been to create single-crystal sheets that have the defect density, composition, and crystal structure of bulk substrate materials. Oxide sheets must be sufficiently thin to produce relevant values of elastic strain using the range of experimentally available stresses. Thin complex oxide sheets have been created by chemical exfoliation, epitaxial lift-off, and lithographic techniques.[16–20] The first two methods, however, face significant experimental challenges. Chemical exfoliation often results in an ensemble of micron-scale sheets rather than a thin layer with large lateral extent.[18] Similarly, sheets created via epitaxial lift-off can face a two-dimensional mechanical constraint imposed by the rigid substrate to which they are transferred.[16]

We focus here on STO sheets fabricated using a lithographic approach starting from a STO single crystal. As shown in Fig. 1(a) the structure consists of a submicron-thick STO sheet with a non-stoichiometric silicon nitride (SiN) layer deposited on each of its planar faces. STO sheets with thickness $t_{STO}$ = 500 nm and area of 10 × 10 µm$^2$ were fabricated from an STO single crystal using focused ion beam (FIB) lithography. Freestanding BTO sheets have been shown to possess bulk like properties, implying they are free of ion-induced strain.[21] However, ion beam damage effects can include Ga ion implantation and the creation of amorphous surface layers that encapsulate and distort nanoscale features. The milling procedure used here was designed to limit the introduction of defects by reducing the milling current for successively finer features ending with a 300 pA beam. Previous studies have shown that annealing and chemical etching can result in a significant reduction in ion-beam-induced damage.[22,23] The removal of ion beam damage in the present study is simplified because we have employed a milling geometry in which the ion beam propagated in a direction within the plane of the large surface of the sheet, limiting the ion propagation into the sheet to a distance approximately equal to the lateral straggle length of approximately 10 nm.[24] The sample was annealed at 1000 ºC for 50 h in order to reduce the distortion of the sheet due to ion-beam induced structural changes, following a procedure described in ref. 19. Previous studies have Indicated under these conditions, the strain in the STO sheet due to milling is on the order of $6 \times 10^{-4}$.[19,25] FIB yielded STO sheets that have the same x-ray nanobeam diffraction rocking curve widths as the bulk unprocessed STO crystal.[19] A scanning electron microscope image of an STO sheet before the deposition of the SiN layers is shown in Fig. 1(b).

The design of the SiN/STO/SiN structure employed a mechanical model based on the balance of forces applied to the sheet, shown schematically in Fig. 1(a).[26] The mechanical calculation was simplified by assuming that the thickness and stress of both SiN layers are equal. Effects arising from edges can be neglected in the central regions of the sheets because the sheets have a large width-to-thickness ratio on the order of 20. Mechanical calculations show that edge effects on the stress distribution are minimal in regions further than approximately one sheet thickness from the edge.[27,28] The SiN sheets apply a biaxial tensile force per unit length $P$, corresponding to a biaxial compressive stress-thickness product equal to $-P/2$ in each of the stressor layers. Each SiN stressor layer has thickness $t_{SiN}/2$. We define a Cartesian coordinate system in which the axes $x$ and $y$ are in the plane of the STO sheet and $x$ is along the bulk substrate surface normal. The non-zero elements of the stress tensor within the sheet are $\sigma_{11} = \sigma_{22} = \sigma_{STO}$, so that the stress applied by nitride layers has a magnitude $\sigma_{STO} = P/t_{STO}$. The fractional change of the interplanar spacing, $d$, along $x$ is $\varepsilon_{STO} = P/(M_{STO} \, t_{STO})$, where $M_{STO} = 395$ GPa is the biaxial modulus for STO (100). [26,29]

The magnitude of the stress-thickness product in the SiN stressors depends on their deposition conditions, thickness, thermal history, and stoichiometry via small differences in the Si:N ratio and in the incorporation of hydrogen during deposition.[30–32] The stress-thickness product in the SiN layers considered here was $P = 28$ GPa nm, measured by evaluating the curvature introduced by a SiN layer deposited on one side of a Si wafer.[33] The mechanical model predicts a strain of $1.4 \times 10^{-4}$ in the STO sheet under these conditions.

The SiN stressor layers were deposited by plasma-enhanced chemical vapor deposition (PT-70, Plasma-Therm) at a sample temperature of 250 °C. The $N_2O$, 2% $SiH_4$ in $N_2$, and 5% $NH_3$ in $N_2$ flow rates were 420, 500, and 80 sccm, respectively, with total pressure of 850 mTorr. The plasma was produced with a power of 36 W at 13.56 MHz and a bias of -28.6 V. The deposition yielded a 270 nm-thick SiN layer conforming to the complex geometry of the STO sheet. Figs. 1(c) and (d) show side and perspective views of the STO sheet after the deposition of the SiN stressors.

The strain introduced in the STO sheets by the SiN stressors was measured using x-ray nanobeam diffraction at station 2-ID-D of the Advanced Photon Source. Fig. 2(a) shows a schematic of the measurement, including definitions of the incident and diffracted x-ray beam angles. The photon energy was 10.1 keV, selected with a (111) Si monochromator. A zone plate with 160 μm diameter and 100 nm-wide outermost zone was used to create a focused beam with a full-width-at-half-maximum diameter of 200 nm. Diffracted x-rays were detected in a horizontal scattering geometry using an x-ray charge-coupled device (Princeton Quad-RO, Princeton Instruments) with 24 μm pixels located 1 m from the sample. The focused beam had a convergence angle of 0.07°, which makes it impossible to define a unique angle for the incident and diffracted beams. We instead define the effective incident angle, $\theta$, to be the angle between the central axis of the focused x-ray beam and the (200) planes in the STO substrate. The angle $2\theta$ refers to the projected angle in the horizontal plane between the center of the focused x-ray beam and the location of each detector pixel.

X-ray diffraction results were compared to an optical simulation in order to allow the strain and the tilt of the lattice to be determined separately. The simulation propagates the incident x-ray beam from the zone plate to the sample, computes the amplitude and phase of the diffracted beam, and propagates the diffracted beam from the sample to the detector.[34–36] A calculation of the reflected amplitude using the kinematic approximation accurately reproduces the x-ray diffraction patterns of the sheets. The diffraction patterns of the thick bulk STO reference sample, however are not adequately described using kinematic scattering and were instead calculated using the Darwin dynamical theory. Dynamical effects were incorporated into the simulation by assuming that the bulk STO crystal consisted of a perfect stack of STO unit cells with infinite thickness.[36]

Bulk STO was used as a reference for the orientation and spacing of the (200) planes in the STO sheet. Measured diffraction patterns and dynamical diffraction simulations for the bulk STO substrate are shown in Figs. 2(b) and 2(c), respectively. The experimental diffraction pattern from an STO sheet and the corresponding kinematic diffraction simulations are shown in Figs. 2(d) and 2(e).

The angular widths of the high-intensity features in the diffractions patterns in Fig. 2 are significantly different for the STO sheet and bulk STO. For the sheets, for which the kinematic approximation is appropriate, the angular width is inversely proportional to the number of atomic planes illuminated by the x-ray beam, leading a comparatively broad $9 \times 10^{-3}$ degree-wide intensity distribution observed in both in both the experiment and simulation, Figs. 2(d) and 2(e). The measured angular width of the dynamical reflection of the bulk substrate is $4 \times 10^{-3}$ deg. This width includes contributions from the Darwin reflectivity, which has angular width $3 \times 10^{-3}$ deg,[36,37] and the $4 \times 10^{-3}$ deg variation in the Bragg angle introduced by the 2 eV energy bandwidth of the x-ray radiation.

The strain and tilt were separately measured by collecting diffraction patterns over a range of incident angles greater than the zone plate divergence. The distribution of intensity along the vertical direction of the diffraction patterns does not depend on the STO lattice parameter or on the tilt in the $\theta$ angular direction and thus can be vertically integrated to summarize diffraction patterns obtained at different incident angles. The simulated vertically integrated intensity is shown in Fig. 3(a) for several values of the incidence angle $\theta$. Shifts of the center of the intensity along the $\Delta 2\theta$ direction of Fig. 3(a) arise from a change in the lattice parameter caused by elastic strain. Shifts in the $\Delta\theta$ direction arise from changes in the orientation of the STO crystal, which can occur without a change in lattice parameter.

Simulations of several combinations of strain and tilt were conducted to illustrate how these quantities can be separately extracted from experimental diffraction patterns. Fig. 3(b) shows a simulated diffraction pattern for an STO sheet with zero strain and an orientation offset by 0.03°. The region of high intensity in Fig. 3(b) is shifted exclusively in the $\Delta\theta$ direction with respect to the untilted crystal. Fig. 3(c) shows a complementary simulation in which the STO sheet has an in-plane strain of -0.2% with zero tilt, resulting in a shift along the $\Delta 2\theta$ and $\Delta\theta$ directions. STO sheets with (i) zero tilt and zero strain, (ii) non-zero tilt, and (iii) non-zero strain are illustrated in Fig. 3(d).

The strain induced by the SiN was experimentally evaluated by measuring $\Delta 2\theta$ using x-ray rocking curves acquired within a 9 μm wide × 4 μm high area of x-ray beam positions. Fewer measurements were made of the STO sheets without SiN stressor layers due to time constraints during the x-ray nanobeam diffraction experiment. Histograms of strain measured at many locations within these areas of STO sheets with and without SiN stressor layers are shown in Fig. 4(a). The strain in each case was computed using the bulk STO (200) reflection as a reference. The histogram of strain in the STO sheets without SiN stressor layers is shifted by $6 \times 10^{-4}$ with respect to bulk STO.[21] Strain of a similar magnitude was observed in previous studies of FIB-patterned STO, and has a magnitude consistent with the formation of point defects during ion-beam milling.[19]

The strain measured from the STO sheet with the stressors has a different distribution than the sheet without the stressor, as shown in Fig. 4(a). The most commonly observed strain was shifted to a higher value in the sheet with the nitride stressors. We have evaluated the strain using the most commonly occurring value of strain, or mode, rather than the mean because of the non-Gaussian distribution of the measured values and the small number of points at which the strain was measured in bare STO sheets. Small changes in the interplanar spacing on the (200) planes may arise from point-to-point variation in the SiN composition or from irregularities at the SiN-STO interfaces. The difference between the modes of the strain distributions shown in Fig. 4(a) is $1.3 \times 10^{-4}$. This measured strain agrees with the mechanical model prediction of $1.4 \times 10^{-4}$.

This observation of elastic strain sharing in STO sheets also points to the potential use of this approach under a wide range of conditions that can be reached using the same materials. Fig. 4(b) illustrates a range of values of strain that can be produced with strain sharing by modifying the thickness of the STO sheet and stress-thickness product. The strain predicted using the mechanical model is shown for stress-thickness products $P =$ 15, 28, and 45 GPa nm. These values of $P$ are within the range of values available in SiN deposition.[32] The strain measured for the STO sheet probed in this experiment is indicated as an asterisk at a thickness of 500 nm in Fig. 4(b) and falls near the black line representing the stress-thickness product applied in this experiment.

Elastic strain sharing in lithographic structures is a widely applicable strategy for producing elastically strained complex oxides. STO sheets can be elastically strained through the deposition of SiN thin films to induce strain with a magnitude sufficiently large to modify physical properties relevant to magnetism, ferroelectricity, and ionic transport. Selecting the stress-thickness product in the stressors and the thicknesses of the sheets provides the opportunity to precisely choose the elastic strain. This control is achieved without changes in composition or increases in dislocation density, as evidenced by the observed rocking curve widths.[19] The approach can be extended to sheets of larger lateral size and can provide a route to complex oxides with improved properties for a range of technological applications in electronics and optics.

**Acknowledgements**

This work was supported by the University of Wisconsin Materials Research Science and Engineering Center, NSF grant nos. DMR-1121288 and DMR-1720415. The development of x-ray nanodiffraction analysis methods was supported by the U.S. DOE, Basic Energy Sciences, Materials Sciences and Engineering, contract no. DE-FG02-04ER46147. J.A.T. acknowledges support from the National Science Foundation Graduate Research Fellowship Program, grant no. DGE-1256259. Use of the Advanced Photon Source was supported by the U.S. Department of Energy, Office of Science, Office of Basic Energy Sciences, under Contract No. DE-AC02-06CH11357.

**Submitted**, (2017).

Figure 1. (a) Schematic of the STO sheet with SiN stressor layers. Scanning election microscopy images of a 500-nm-thick STO sheet (b) without (c-d) with a 270-nm-thick SiN layer deposited on each face. The large faces of the sheet have area $10 \times 10 \ \mu m^2$.

Figure 2. (a) Schematic of the x-ray scattering geometry showing the definitions of the angles $\theta$ and $2\theta$. (b) Diffraction patterns of the (200) STO bulk Bragg reflection. (c) Simulated dynamical diffraction patterns of the (200) bulk Bragg reflection. (d) Measured and (e) simulated kinematic diffraction patterns of the (200) reflection of the STO sheet.

Figure 3. Simulated STO rocking curves for (a) unstrained STO with zero tilt, (b) unstrained STO with a 0.1° tilt, and (c) -0.2% strained STO with zero tilt. The dotted lines and crosses are a guide to the eye. (d) Schematic of tilt and strain in the STO sheets with lattice spacing $d$.

Figure 4. (a) Histogram of strain measured from the (200) rocking curves in the STO sheets without a SiN layer (blue cross hatched) and with a SiN layer (red cross hatched). The change in lattice parameter of the bare STO sheet arises from the FIB processing. The increase in lattice parameter after SiN deposition agrees with the mechanical prediction for an elastically strained sheet. (b) Predicted strain in STO sheets as a function of thickness of the STO sheet for different stress-thickness product. The black line is the experimental stress, and the black asterisk represents the experimentally measured stress and strain.

(a)

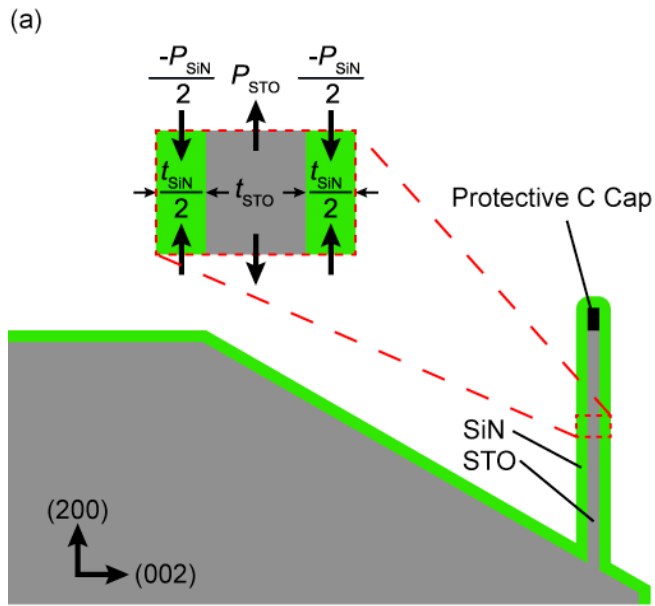

$\frac{-P_{SiN}}{2}$   $P_{STO}$   $\frac{-P_{SiN}}{2}$

$t_{SiN}$   $t_{STO}$   $t_{SiN}$
$\frac{}{2}$           $\frac{}{2}$

Protective C Cap

SiN
STO

(200)
(002)

(b)

5 μm

(c)

1 μm

(d)

5 μm

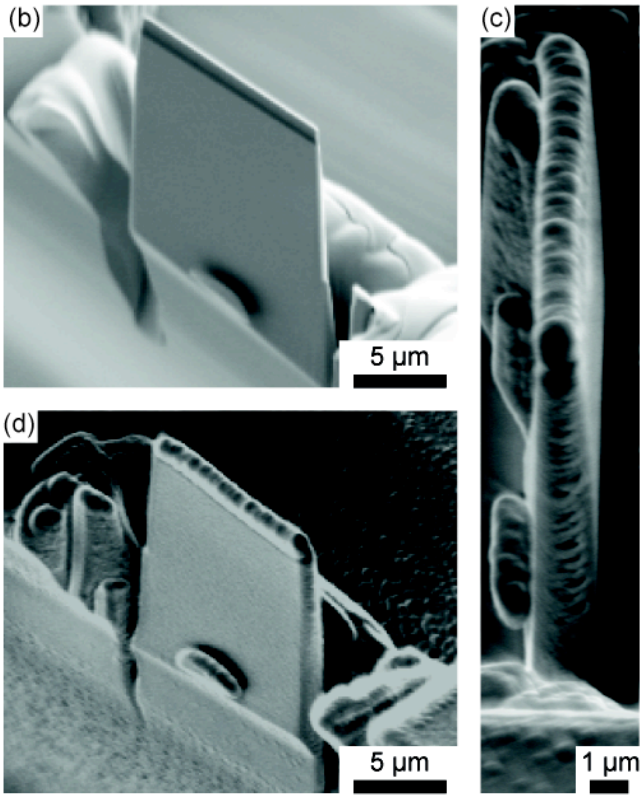

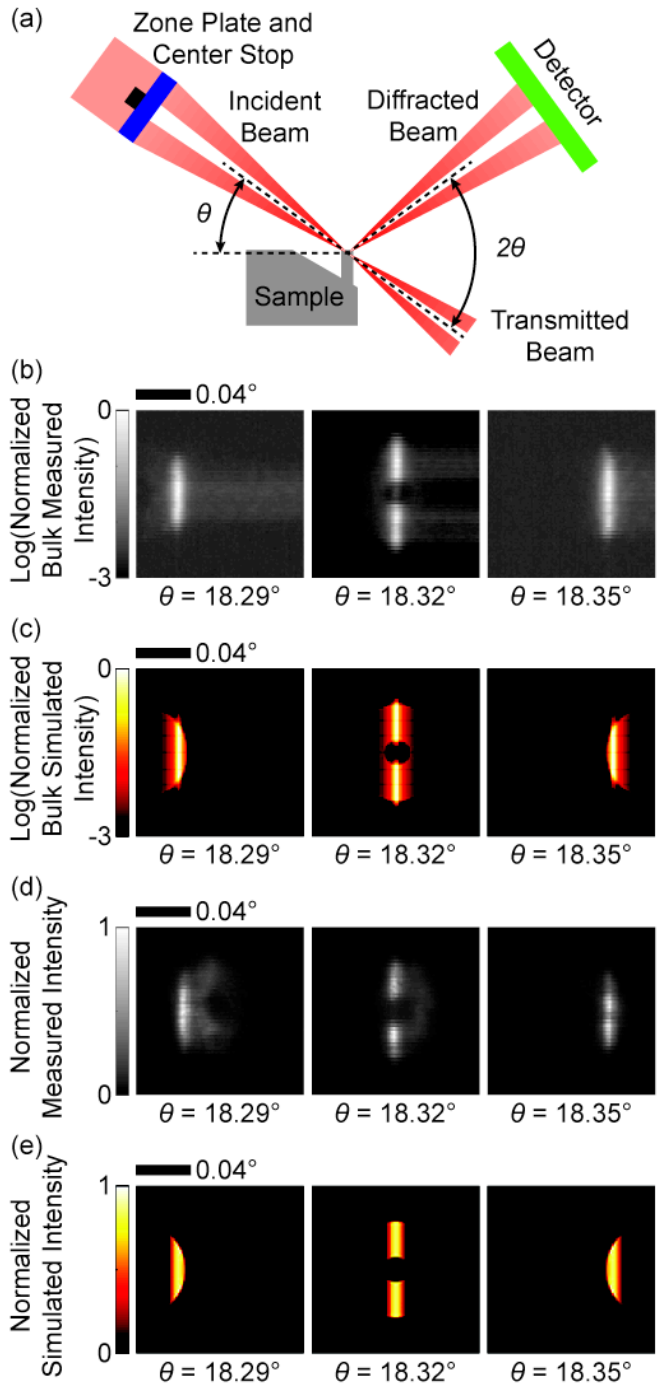

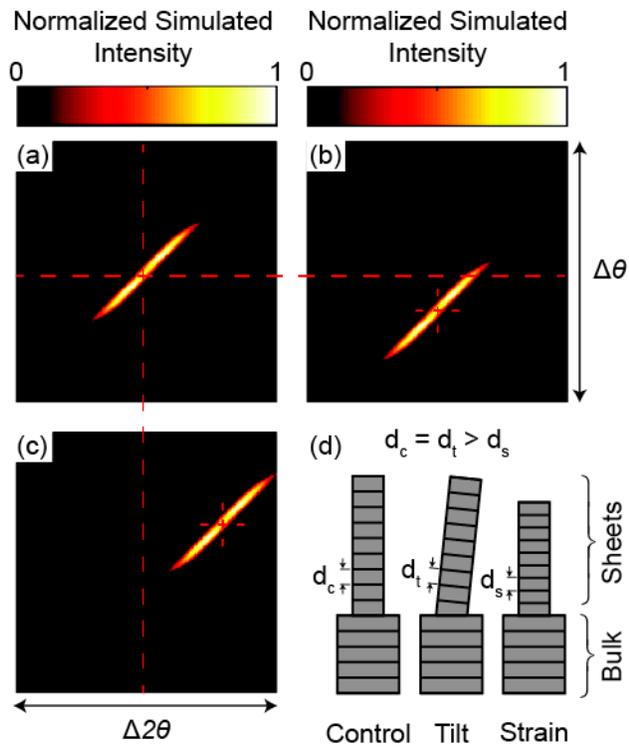

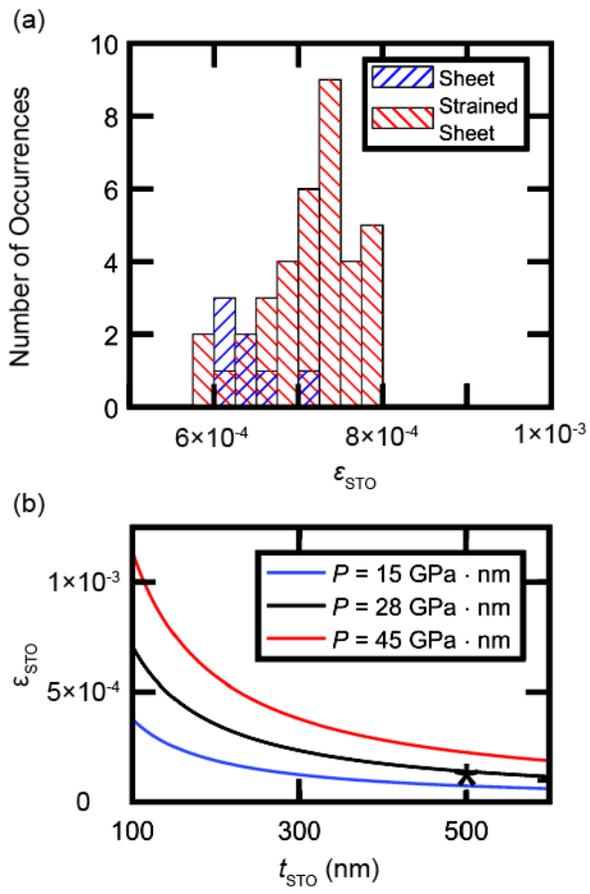